\documentclass[sigconf]{acmart}

\usepackage{tikz}
\usetikzlibrary{patterns,shapes,arrows}
\usetikzlibrary{positioning,fit,calc}
\usetikzlibrary{arrows,automata}
\usepackage{pgfplots}
\pgfplotsset{compat=1.14}
\usepackage{algorithm}
\usepackage{algorithmic}
\usepackage{amsmath}
\DeclareMathOperator*{\argmax}{argmax}

\usepackage{microtype}
\usepackage{multirow}

\setcopyright{none} 
\acmConference[Anonymous Submission to ACM CCSW 2020]{ACM Cloud Computing Security Workshop}{Due 31 July 2020}{TBD}
\acmYear{2020}

\tikzset{ablock/.style={draw, thick, text width=4cm, minimum height=1cm, align=center, node distance=.2cm},
         aempty/.style={rectangle,   text width=4cm, minimum height=1cm, align=center, node distance=.2cm},
         apoint/.style={rectangle,   text width=4cm, minimum height=1cm, align=center, node distance=.2cm},
         end/.style   ={rectangle,   text width=.5cm,  minimum height=1cm, align=center, node distance=.2cm},
         acirc/.style={draw, circle, thick, text width=2cm, minimum height=1cm, align=center, node distance=.2cm},
         dot/.style={node distance=.2cm},
         line/.style = {-latex}
}

\newenvironment{mylist}
  {\list{}{%
    \leftmargin=2.4em
    \topsep=2ex
    \parsep=\parskip
    \listparindent=\parindent
    \itemindent=\parindent
  }\item\relax}
  {\endlist}

\author{Blake Anderson}
\affiliation{\institution{Cisco}}
\email{blake.anderson@cisco.com}

\author{David McGrew}
\affiliation{\institution{Cisco}}
\email{mcgrew@cisco.com}

\begin{document}
\title{Accurate TLS Fingerprinting using Destination Context and Knowledge Bases}

\begin{abstract}

Network fingerprinting is used to identify applications, provide insight into network traffic, and detect malicious activity. With the broad adoption of TLS, traditional fingerprinting techniques that rely on clear-text data are no longer viable. TLS-specific techniques have been introduced that create a fingerprint string from carefully selected data features in the \texttt{client\_hello} to facilitate process identification before data is exchanged. Unfortunately, this approach fails in practice because hundreds of processes can map to the same fingerprint string. We solve this problem by presenting a TLS fingerprinting system that makes use of the destination address, port, and server name in addition to a carefully constructed fingerprint string. The destination context is used to disambiguate the set of processes that match a fingerprint string by applying a weighted na\"{i}ve Bayes classifier, resulting in far greater performance.

Our methods are made possible by a data fusion system that continuously collects and fuses host and network data, building up-to-date fingerprint knowledge bases that correlate TLS fingerprint strings, processes, and destinations for 50+ million real-world sessions each day.  Using data collected from two geographically distinct sites and a malware analysis sandbox, we demonstrate that our solution can achieve an $F_1$ score of greater than 0.99 for process identification and high efficacy malware detection with 99.9\% precision and 88.7\% recall. We provide specific results for the set of most common processes and a set of cloud orchestration tools. In the case of no exact fingerprint string matches, we demonstrate that our system can accommodate approximate fingerprint string matching with an $F_1$ score of 0.90. Finally, we have released an open source tool, mercury \cite{mercury}, that implements the proposed techniques and provide weekly updates to an open source TLS fingerprint knowledge base to assist reproducibility of our work.

\end{abstract}

\ccsdesc[500]{Security and privacy~Network security}
\ccsdesc[300]{Security and privacy~Malware and its mitigation}

\keywords{TLS Fingerprinting; Process Identification; Malware}

\maketitle

\section{Introduction}

Process identification from network traffic aids many use cases including network segmentation, malware detection, and vulnerable application detection. The HTTP \texttt{User-Agent} \cite{http11_semantics_content} has been used as a proxy for process identification, but with the increasing use of Transport Layer Security (TLS) \cite{tls12,tls13}, methods that rely on clear-text data have become obsolete. Existing solutions have been proposed for identifying processes that have initiated TLS connections \cite{taylor16appscanner,van2020flowprint}, but these solutions must observe complete sessions making them unsuitable for real-time enforcement.

TLS fingerprinting has been proposed as a technique to enable real-time enforcement by providing the initiating process's identity after observing the client's initial TLS packet. Traditional TLS fingerprinting extracts metadata presented in the TLS \texttt{client\_hello} and generates a fingerprint string using a pre-defined schema. These techniques are relevant for all versions of the TLS protocol, including TLS 1.3 \cite{tls13} where all needed data features are still presented unencrypted. Given a fingerprint string, TLS fingerprinting then maps that string to a process by using a dictionary of known fingerprint-to-process mappings. Unfortunately, TLS fingerprint strings are often more indicative of a TLS library than they are of a specific process, with fingerprint strings often mapping to tens or hundreds of unique processes.

This limitation can be seen by analyzing publicly available malware fingerprint feeds. Abuse.ch, which cautions that their feed has ``not been tested against known good traffic yet and may cause a significant amount of FPs", provides a list of 70 JA3 \cite{salesforce17ja} hashes used by malware. We reverse engineered 68 of the hashes and mapped them to our data format. 59 of the indicators were more strongly associated with benign processes, such as Internet Explorer, Python, and Java, than they were with malware. 12 of the indicators led to 1 million or more false positives during a 10-day period for one of our testing sites. While the TLS fingerprint string taken by itself is often a poor indicator, additional contextual information can help to increase performance.

In this paper, we generalize TLS fingerprinting by incorporating contextual information contained within the \texttt{client\_hello} packet. Our approach uses the destination IP address, port, and \texttt{server\_name} value (if available) to disambiguate potential processes. We define equivalence classes for the destination features to help generalize to unseen destination values. As an example, the classification system uses both the IP address and the autonomous system of the IP address. We combine the features using a simple weighted na\"{i}ve Bayes classifier, which relies on probability estimates provided by our fingerprint knowledge base. We show that our approach of simultaneously considering the TLS fingerprint string and the destination information is a significantly improvement compared to systems based solely on the fingerprint string or the destination information.

An underlying assumption of TLS fingerprinting is that there exists a well-curated database that maps TLS fingerprint strings to process or library names. This is especially true for our work, where the na\"{i}ve Bayes algorithm requires a knowledge base that provides prevalence information for each process associated with a fingerprint string, along with counts for each destination feature that a given process visited while using a specific TLS fingerprint string. We built a system that continuously fuses real-world host and network data, which we use to build a TLS knowledge base that reflects the most recent TLS usage on the monitored networks based on billions of connections. Furthermore, the automated nature of our knowledge base generation ensures that our system stays current with destination information that frequently changes.

To demonstrate the efficacy of the proposed solution, the classifier is trained on data collected from a site in GMT+19 during the month of May 2020 and applied to data collected during the first ten days of June 2020 from the same site as well as a site in GMT+8. We provide the process identification $F_1$ score, as well as precision/recall results highlighting how the classifier performs on the top applications and cloud orchestration and processing tools. Malware's use of TLS has been well-documented \cite{anderson16deciphering}, with many malware families prone to using standard libraries as discussed above. We use the results of a malware analysis sandbox to demonstrate that our techniques are able to achieve 99.9\% precision and 88.7\% recall for the malware detection task.

Operational concerns or parameters are often ignored in the construction and application of TLS fingerprint databases. TLS fingerprint strings and the way applications use TLS is constantly evolving \cite{anderson2019beyond,kotzias2018coming}, leading to a steady introduction of new fingerprint strings. Accommodating approximate matching helps manage unseen fingerprint strings, and we show that an approximate matching scheme based on edit distance can still achieve an $F_1$ score up to 0.902 on sessions with an unknown fingerprint string. We also demonstrate the importance of a well-tuned knowledge base by looking at the classifier's performance when it is not kept up to date. A knowledge base in our format that captures all process and destination information for hundreds of billions of sessions can be up to several hundred megabytes, reducing the efficiency of online classification. We show that incorporating data that is older than one month reduces classification performance and unnecessarily increases the size of the database.

To support this work, we developed an open source C/C++ tool, mercury \cite{mercury}, that uses Linux's AF\_PACKET TPACKETv3 zero-copy shared memory ring buffers to collect and classify data on network links with capacities of 30Gbps+. We also released a python-based implementation to facilitate rapid prototyping. We are committed to releasing up-to-date TLS fingerprint knowledge bases, which we have currently done weekly during the first seven months of 2020. Some information has been removed from the open source fingerprint knowledge base relative to the internal knowledge bases, and we highlight the difference in expected performance in Section \ref{sec:reproducibility}.

Our novel contributions include:
\begin{itemize}
  \item The introduction of a TLS fingerprinting system that incorporates destination information and accommodates approximate matching to provide state-of-the-art process and malware identification results.
  \item A study of the impact of different knowledge base configuration options on the classifier's performance.
  \item An open source tool that implements all presented techniques along with a TLS fingerprint knowledge base that is updated weekly and currently has over 8,400 TLS fingerprint strings with detailed process and destination information.
\end{itemize}


\section{TLS Fingerprinting}
\label{sec:tls-fingerprinting}

\begin{figure}
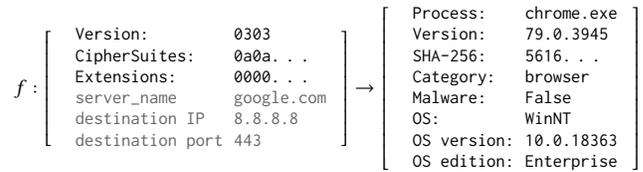

  \centering
  \footnotesize
  $f: \left[
    \texttt{
    \bgroup
    \setlength{\tabcolsep}{2pt}
    \begin{tabular}{ll}
    Version:      & 0303 \\
    CipherSuites: & 0a0a\ldots \\
    Extensions:   & 0000\ldots \\
    \textcolor{black!60}{server\_name} & \textcolor{black!60}{google.com} \\
    \textcolor{black!60}{destination IP} & \textcolor{black!60}{8.8.8.8} \\
    \textcolor{black!60}{destination port} & \textcolor{black!60}{443} \\
    \end{tabular}
    \egroup
    }
  \right] \rightarrow \left[
    \texttt{
    \bgroup
    \setlength{\tabcolsep}{2pt}
  \begin{tabular}{ll}
    Process:             & chrome.exe \\
    Version:             & 79.0.3945 \\
    SHA-256:             & 5616\ldots \\
    Category:            & browser \\
    Malware:             & False \\
    OS:                  & WinNT \\
    OS version:           & 10.0.18363 \\
    OS edition:           & Enterprise
  \end{tabular}
  \egroup
  }
       \right]$
  \caption{Classical TLS fingerprinting aims to map parameters extracted from the \texttt{client\_hello} to a set of informative labels such as the process name or operating system. Gray features are only used in generalized TLS fingerprinting.}
  \label{fig:tls-fp-example}
\end{figure}

Transport Layer Security (TLS) \cite{tls12,tls13} is the most popular protocol to secure communications over the Internet. A client begins a TLS session by sending a \texttt{client\_hello} message, which contains a TLS version, a list of supported cipher suites ordered by preference, and a list of extensions that provide additional context, e.g., the \texttt{server\_name} extension provides the DNS hostname so that front-end servers can route connections without having to perform decryption \cite{rfc6066}. The server then responds with a \texttt{server\_hello} message that selects a set of cryptographic parameters offered by the client and a \texttt{certificate} message proving the server's identity. The server then initiates the key exchange, after which the client and server send \texttt{finished} messages and begin exchanging encrypted data.

TLS fingerprinting operates on the initial \texttt{client\_hello} message, allowing for real-time policy enforcement before the TLS handshake completes or encrypted data is exchanged. Figure \ref{fig:tls-fp-example} shows a simple example of an idealized TLS fingerprinting system, where the goal is to learn some function, $f$, that maps parameters offered in the \texttt{client\_hello} to a set of informative labels such as the process name or operating system. In this work, we focus on identifying the process name.

Several goals are important for the generation of a TLS fingerprint string. The fingerprint string format should be unambiguous and reversible. It should provide the most discriminating power possible, taking advantage of every informative data feature. It should be able to accommodate complex patterns such as TLS GREASE \cite{grease}. Lastly, it should be computationally inexpensive, and robust against Denial of Service (DoS) attacks that aim to exhaust system resources. Our system uses the following fingerprint schema:

\begin{mylist}
\texttt{(version)(cipher suites)((ext$_1$)(ext$_2$)...)}
\end{mylist}

\noindent where each field is a hex string corresponding to the bytes observed in the \texttt{client\_hello} to facilitate reversibility, i.e., it is possible to determine substrings that were in the packet from the fingerprint string. To help ensure discriminating power, all cipher suite and extension orderings are maintained, and \texttt{((ext$_1$)(ext$_2$)...)} includes the data for 21 extensions along with all type codes. The data associated with session-specific extensions is omitted, e.g., \texttt{server\_name} and \texttt{key\_share}, and data associated with client-specific parameters is kept, e.g., \texttt{supported\_groups}, \texttt{supported\_versions}, and \texttt{compress\_certificate}. A full list of extensions that retain their data in the fingerprint schema is given in Appendix \ref{app:tls-extensions}. Along with normalizing session-specific extension data, GREASE \cite{grease} cipher suites, extension types, and extension data values are normalized to the value \texttt{0a0a}, but their ordering is preserved. Finally, we avoid any cryptographic computations on the fingerprint string itself, such as computing an MD5 hash, for computational efficiency.

Given a TLS fingerprint string, traditional fingerprinting systems would return a single process or set of processes that have been observed using that fingerprint string as defined in a fingerprint database. Unfortunately, TLS fingerprint strings often map to many processes due to those processes using the same underlying TLS library. For example, in the data we collected during the month of May 2020, the median number of unique process names for the top-100 most prevalent fingerprint strings was 24.5. To provide actionable intelligence, a TLS fingerprinting system needs significantly more specificity than current approaches provide.

\section{Generalized TLS Fingerprinting}
\label{sec:generalized-tls-fingerprinting}

To overcome the limitations of previous systems, we put forward an approach to generalize TLS fingerprinting as explained in this section. Our approach is centered around the construction of a TLS fingerprint knowledge base that relies on continuous, large-scale data collection, curation, and fusion. We also propose an approximate matching scheme to accommodate the introduction of new fingerprint strings for completeness. To further improve the robustness of our system, we introduce equivalence classes on destination features to better handle unseen destination/process combinations. Finally, given a match in our knowledge base, we use a weighted na\"{i}ve Bayes classifier to provide the most probable process name given the list of potential processes and their destinations.

\subsection{Knowledge Base Generation}
\label{sec:knowledge-base}

\begin{figure}
  \centering
  \includegraphics[scale=3.1]{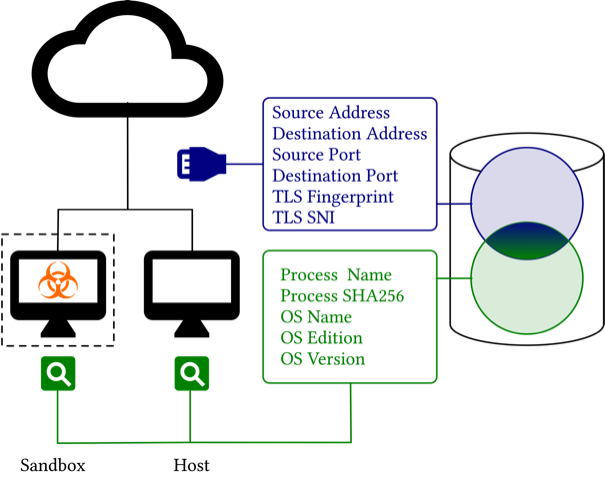}
  \caption{Data fusion system that correlates network and host logs in order to attribute network sessions to processes.}
  \label{fig:data-fusion-pipeline}
\end{figure}

Traditional TLS fingerprint databases provide mappings from TLS fingerprint strings to processes, libraries, or operating systems, but lack the context needed to disambiguate the results. The TLS fingerprint knowledge base provides this context by associating each TLS fingerprint string with a list of processes observed using it, along with destination and operating system information. We have built a data fusion system as shown in Figure \ref{fig:data-fusion-pipeline} to collect this data. For the host data, we use records sent by the AnyConnect Network Visibility Module (NVM) \cite{anyconnect}, which contain the network 5-tuple, an event start timestamp, the name of the communicating process, the SHA-256 hash of the process executable, and the host's operating system. For the network data, we use our custom open source tool, mercury \cite{mercury}, which reports the TLS fingerprint string along with the network 5-tuple and an event start timestamp.

The host and network data are joined daily by the network 5-tuple and event start timestamps. If there are network 5-tuple collisions, the records with the minimal timestamp delta are joined. We discard records with timestamp deltas greater than 5 seconds, which resulted in less than a 0.1\% data reduction. The joined records contain the TLS fingerprint string, destination information such as the IP address and \texttt{server\_name} extension value, and process attribution information.

The fused records are then used to condition a knowledge base. Records are grouped by their TLS fingerprint string, collecting a list of all associated processes. For each process, we record the total number of sessions and all destinations with associated counts. Each destination is represented as a 3-tuple comprised of the IP address, port, and \texttt{server\_name} value (if present). Separate knowledge bases are generated for each day's traffic, and then merged to create an operational knowledge base. We use this flexibility to discard older data as described in Section \ref{sec:operationalizing}.

In addition to real-world host and network data, we also generate knowledge bases from the artifacts of a malware analysis sandbox. The joining procedure is similar but relies on packet captures and analysis files to create the joined records.

\subsection{Approximate Matching}
\label{sec:approx-matching-desc}

\begin{figure}
  \centering
  \begin{center}
    \resizebox{!}{.25\textheight}{
    \begin{tikzpicture}                                                                                                                                      
      \node[aempty]                             (a)  {\begin{tabular}{c}  packets \end{tabular}};
      \node[ablock, below= of a, fill=gray!10]  (ab) {\begin{tabular}{c} protocol \\ identification \end{tabular}};
      \node[ablock, below= of ab, fill=gray!10] (b)  {\begin{tabular}{c} fingerprint \\ extraction \end{tabular}};
      \node[ablock, below= of b, fill=gray!10]  (bb)  {\begin{tabular}{c} substring \\ normalization \end{tabular}};
      \node[ablock, below= of bb, fill=gray!10] (d)  {\begin{tabular}{c} exact \\ matcher \end{tabular} };
      \node[ablock, below= of d, fill=gray!10]  (f)  {\begin{tabular}{c} approximate \\ matcher \end{tabular}};
  
      \draw[->] (a) -- (ab);
      \draw[->] (ab) -- (b);
      \draw[->] (b) -- (bb);
      \draw[->] (bb) -- (d);
      \draw[->] (d) -- (f);

      \draw [->, thick] ($ (d.east) $) -- ($ (d.east) + (1cm,0) $);
      \draw [->, thick] ($ (f.east) $) -- ($ (f.east) + (1cm,0) $);

      \draw [->, thick, dashed]
         ($ (f.west) $)
      -- ($ (f.west) + (-1cm,0) $)
      -- ($ (d.west) + (-1cm,0) $)
      -- ($ (d.west) $);
    \end{tikzpicture}
  }
  \end{center}
  \caption{Control flow for online matching.}
  \label{fig:fingerprint-control-flow}
\end{figure}
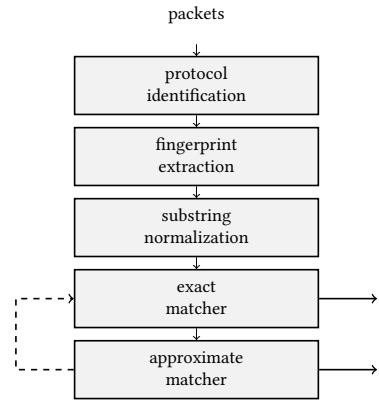

TLS fingerprint strings are constantly evolving. Kotzias et al. \cite{kotzias2018coming} demonstrated this evolution and how the disclosure of security flaws in TLS impacts the TLS versions, cipher suites, and extensions clients offer. Our full knowledge base, collected between July 2019 to June 2020, contains over 8,000 fingerprint strings with process information, but we still consistently add 10-20 new fingerprint strings with process information per day.

Similar to previous work \cite{frolov2019censorship}, we dealt with this issue by implementing a fingerprint string similarity metric based on the Levenshtein distance, which measures the number of cipher suite and extension insertions, deletions, or substitutions that are needed to transform one fingerprint string into another. When we see a fingerprint string that is missing from our database during online analysis, we compute the Levenshtein distance between all known fingerprint strings and select the fingerprint string with the minimal distance. The fingerprint string's prevalence is used to break any ties. Results specific to approximate matching are given in Section \ref{sec:approx-matching}

Figure \ref{fig:fingerprint-control-flow} demonstrates the system's control flow from observing a packet on the wire to reporting the appropriate entry in the knowledge base. We first perform protocol identification using a bit mask over the first 8 bytes of application data, matching known TLS versions and the \texttt{client\_hello}'s record and message identifier. We then extract the fingerprint string conforming to the schema described in Section \ref{sec:tls-fingerprinting}, and normalize session-specific extension data and GREASE values. We then report the exact match if the fingerprint string is currently in our knowledge base or use the approximate matching technique as described above. While approximate matching is computationally expensive, we store the results in the knowledge base which leads to a low amortized cost.

\subsection{Equivalence Classes}

Similar to the introduction of new fingerprint strings, the destination information associated with a process can change over time. The changes often take the form of a new subdomain in the \texttt{server\_name} or a different IP address within the same public cloud environment. Our classification system described in Section \ref{sec:naive-bayes} uses probabilities conditioned on real-world observations to select the most probable process. If a \texttt{server\_name} has a random component in the subdomain but the domain name remains constant, this would result in a zero probability despite the \texttt{server\_name} being obviously related to the known process.

We solve the problem of unseen destination values by introducing equivalence classes for the observable features, and we use those equivalence classes in the classifier. Each equivalence relation partitions the set of features into distinct subsets. There may be more than one useful equivalence relation for a feature, since there are multiple ways that addresses, ports, and domain names can be related.  All the addresses within a BGP Autonomous System (AS) \cite{rfc6793} are equivalent in some way, as are the addresses in a corporate offering such as Microsoft Office 365, Azure, AWS, or Cloudflare.

In the current work, we test four equivalence classes. For the \texttt{server\_name} feature, we extract the domain name and the top-level domain using Mozilla's Public Suffix List \cite{publicsuffixlist}. The IP address is mapped to the corresponding BGP AS using MaxMind's GeoLite2 database \cite{maxmind}. The ports are mapped to a known application layer protocol, e.g., 443 $\rightarrow$ HTTPS and 993/995 $\rightarrow$ email, with unknown and ephemeral ports mapping to ``unknown". While a more interesting set of equivalence classes would undoubtedly improve the performance of the classification system, it would not change the mechanics of the classifier and we leave these investigations to future work as discussed in Section \ref{sec:discussion}.

\subsection{Weighted Na\"{i}ve Bayes Model}
\label{sec:naive-bayes}

\bgroup
\def\arraystretch{1.1}
\begin{table}
\center
\begin{tabular}{lr}
  \hline
  \multicolumn{1}{c}{Feature} & \multicolumn{1}{c}{Weight} \\
  \hline
  \texttt{server\_name}                      & 0.97192 \\
  \texttt{server\_name} $\rightarrow$ Domain & 0.16200 \\
  \texttt{server\_name} $\rightarrow$ TLD    & 0.01044 \\
  IP                                         & 0.53294 \\
  IP $\rightarrow$ AS                        & 0.10343 \\  
  Port                                       & 0.00396 \\
  Port $\rightarrow$ Port Class              & 0.00265 \\  
  \hline
\end{tabular}
\caption{Weighted na\"{i}ve Bayes feature weights as found with the information gain ratio.}
\label{table:feature-weights}
\end{table}
\egroup

Given an exact or approximate fingerprint string match and a set of destination features along with their equivalence mappings, our goal is now to select the most probable process from the fingerprint string's list of possible processes.

We formalize the system as follows. Each session is associated with a process $z$, which is not directly observed, as well as the observed destination features and equivalence mappings, $f_1, \ldots, f_n$. $\mathcal{Z}$ denotes the set of processes previously observed using the matched fingerprint string as given by the knowledge base. Our goal is to construct a classifier $c$ that given a TLS fingerprint string, $f_p$, returns the process that maximizes $\mathbf{P}(z \mid f_1, \ldots, f_n)$ for $z \in \mathcal{Z}_{f_p}$.

For interpretability and computational efficiency, we chose the na\"{i}ve Bayes model:
\begin{align} \label{naivebayes1}
  c(f_1, \ldots, f_n) & = \argmax_{z \in \mathcal{Z}_{f_p}} \mathbf{P}(z \mid f_1, \ldots, f_n) \\
   \label{naivebayes2}
  & = \argmax_{z \in \mathcal{Z}_{f_p}} \mathbf{P}(z) \prod_{i=1,n} \mathbf{P}(f_i \mid z)  \\   \label{naivebayes3}
  & = \argmax_{z \in \mathcal{Z}_{f_p}} \log \mathbf{P}(z) + \sum_{i=1,n} \log \mathbf{P}(f_i \mid z). 
\end{align}
Equation \ref{naivebayes1} simply defines the classifier as the function that returns the most probable process given the destination feature set. Equation \ref{naivebayes2} applies Bayes' theorem, removes the irrelevant denominator, and applies the na\"{i}ve Bayes assumption that each observed feature is conditionally independent from all other observed features. Equation \ref{naivebayes3} simplifies the computation and helps to prevent underflow by moving to sums of logarithms.

$\mathbf{P}(f_i \mid z)$ is computed by using the empirical probability estimates provided by the knowledge base. In cases where $\mathbf{P}(f_i \mid z) = 0$, we use a prior probability of $1/t$ where $t$ is the total number of sessions observed using a given fingerprint string in the knowledge base. Computing $\mathbf{P}(f_i \mid z)$ directly from the knowledge base and leveraging a na\"{i}ve Bayes model helps to avoid an unreasonably large number of features that would be needed by other machine learning alternatives such as deep neural networks or support vector machines.

Because the destination features and equivalence mappings provide varying levels of information about the process initiating a TLS session, we opted to modify the na\"{i}ve Bayes algorithm to use a weighted combination of the features as described by Zhang and Sheng \cite{zhang2004learning}. The feature weights are computed using the information gain ratio conditioned on the knowledge base. The weights found with this method using a knowledge base constructed from data collected during May 2020 is given in Table \ref{table:feature-weights}. Both the \texttt{server\_name} and IP address have high weights and are stronger indicators of a process given a fingerprint string relative to the port information, which is heavily biased towards port 443 as shown in Section \ref{sec:data}. Section \ref{sec:feature-importance} provides results when we consider subsets of destination features.

With the feature weights, $w_f$, computed, the new equation to compute the most probable process becomes:
\begin{equation}
  c(f_1, \ldots, f_n) = \argmax_{z \in \mathcal{Z}_{f_p}} \log \mathbf{P}(z) + \sum_{i=1,n} w_{f} \cdot \log \mathbf{P}(f_i \mid z ). 
\end{equation}
Algorithm \ref{alg:naive-bayes} summarizes the procedure to find the most probable process given a fingerprint string from the knowledge base and the destination information from the session.

\begin{algorithm}[t!]
  \caption{Process identification.}
  \label{alg:naive-bayes}
\begin{algorithmic}
\STATE { \textbf{Given}: fingerprint, destination\_features }
\STATE { proc\_list.initialize\_to\_empty\_list() }
\FOR{ $z \in$ fingerprint.processes }
   \STATE { $ q \leftarrow \log \mathbf{P}(z)$  }
   \FOR{ $f \in $ destination\_features }
   \STATE { $ q \leftarrow q + w_{f} \cdot \log \mathbf{P}(f \mid z)$ }
   \FOR{ $\gamma \in$ $f$.eqv\_classes }
         \STATE {$e \leftarrow \gamma(f)$ }
         \STATE { $ q \leftarrow q + w_{\gamma} \cdot \log \mathbf{P}(e \mid z)$ }
      \ENDFOR
   \ENDFOR
   \STATE { proc\_list.append($q$, $z$)    }
\ENDFOR
\STATE { \textbf{return} proc\_list.get\_maximum() }
\end{algorithmic}
\end{algorithm}

\section{Data}
\label{sec:data}

We used mercury \cite{mercury} to collect network data from a site located within GMT+19, referred to as Site 1 (GMT+19) in Section \ref{sec:results}, that also reports host logs from the AnyConnect Network Visibility Module \cite{anyconnect}. The host and network datasets are joined daily as described in Section \ref{sec:knowledge-base}. The data for this paper was collected between 2019-07-01 and 2020-06-10, where the data from June 2020 was exclusively used for testing. Over 70,000 unique hosts generated the data from Site 1.

In total, we observed 9,312 unique fingerprint strings from over 13 billion TLS sessions with associated host data from Site 1 during our monitoring. Table \ref{table:tls-ports} lists the ten most prevalent TLS ports at Site 1. 99.4\% of the TLS sessions used port 443, the typical port for HTTPS. The second most prevalent port, 993, is mainly used for IMAP-over-TLS. This dataset has a large diversity of processes with 22,969 unique process names and 243,736 unique process executable SHA-256s. During the first 10 days in June used for testing, there were 39,768 hosts, 2,320 fingerprint strings, 4,073 process names, 16,474 process executable SHA-256s, and 278,570,891 TLS sessions.

\bgroup
\def\arraystretch{1.1}
\begin{table}[t!]
\center
\begin{tabular}{lr|lr}
  \hline
  \multicolumn{2}{c}{Site 1 (GMT+19)} & \multicolumn{2}{c}{Malware Sandbox}\\
  \multicolumn{1}{c}{Port} & \multicolumn{1}{c}{Sessions} & \multicolumn{1}{c}{Port} & \multicolumn{1}{c}{Sessions}\\
  \hline
  443  & 13,148,433,441 & 443  & 53,082,740 \\
  993  & 40,608,706     & 465  & 701,814 \\
  5228 & 3,430,729      & 9001 & 70,075 \\
  80   & 2,940,342      & 80   & 10,434 \\
  995  & 2,744,560      & 449  & 5,443 \\
  8443 & 2,458,762      & 26   & 4,928 \\
  8080 & 2,418,693      & 8443 & 4,370 \\
  5986 & 2,222,676      & 993  & 3,762 \\
  465  & 1,801,507      & 9002 & 3,690 \\
  5223 & 1,527,906      & 8080 & 3,475 \\
  \hline
\end{tabular}
\caption{Top-10 TLS ports from Site 1 (GMT+19) and the malware analysis sandbox.}
\label{table:tls-ports}
\end{table}
\egroup

To perform additional validation and to test how well the knowledge base generalizes to new locations, we collected the same joined data from a geographically distinct site located in GMT+8, referred to as Site 2 (GMT+8) in Section \ref{sec:results}. This site belonged to the same enterprise; we leave validation on distinct entities for future work as described in Section \ref{sec:discussion}. We collected this data from 2020-06-01 to 2020-06-10, and it was only used for testing. There were 10,175 hosts, 824 fingerprint strings, 1,471 process names, 5,210 process executable SHA-256s, and 33,820,842 TLS sessions.

In terms of the operating systems, nearly 70\% of the data collected from Sites 1 and 2 were MacOS 10.14.6 and Windows 10.0.17134. $\sim$28\% of the data is comprised of other versions of Windows 10 and MacOS 10.15.x. The remaining data is primarily older versions of Windows and MacOS. Only $\sim$0.01\% of the data is Linux-based, mainly Ubuntu 19.04 and 19.10. We further discuss this limitation in Section \ref{sec:discussion}.

Finally, we collected data from a malware analysis sandbox running Windows 7 and 10 between 2019-07-01 and 2020-06-10. Similar to Site 1, the data collected in June 2020 was used exclusively for testing. The full malware dataset has 53,958,368 TLS sessions, 9,348 unique TLS fingerprint strings, and 37,841 unique process executable SHA-256s. As Table \ref{table:tls-ports} shows, the TLS ports for the malware analysis sandbox data were dominated by HTTPS similar to Site 1, with over 98.4\% of the sessions using port 443. There were over 1,200 unique anti-virus signatures associated with 10 or more samples. The most common malware families were Troldesh \cite{troldesh}, Tofsee \cite{tofsee}, Emotet \cite{emotet}, and DarkComet \cite{farinholt2020dark}. The testing data collected in June 2020 had 347 fingerprint strings, 9,117 process executable SHA-256s, and 298,126 TLS sessions.

\section{Results}
\label{sec:results}

Throughout this section, we present results related to process identification, identifying cloud orchestration and processing tools, malware detection, and the importance of destination feature subsets. We compare the performance of the weighted na\"{i}ve Bayes, unweighted na\"{i}ve Bayes, and ``top process" classification methods, where top process ignores the destination features and simply selects the process with the most observations given a TLS fingerprint string. The top process method still takes advantage of the process prevalence information in the knowledge base. We present results using the weighted na\"{i}ve Bayes classification method if not stated otherwise. We further compare our approach to methods that make no use of the fingerprint string, instead relying solely on either the TLS \texttt{server\_name} or the destination IP address.

To summarize overall performance, we use the micro-averaged $F_1$ score where the label set is either a process name or a process family name. The process name label set has been normalized so that processes appearing on different platforms, e.g., MacOS and WinNT, map to the same label. For example, \texttt{chrome.exe} on WinNT and \texttt{google chrome} on MacOS both map to \texttt{chrome}. The process family name labels group sets of process that share an underlying architecture and purpose. The two largest and most diverse process families are Microsoft Office, including processes like \texttt{excel.exe}, \texttt{outlook.exe}, and \texttt{word.exe}, and Chromium-based web browsers, including processes like \texttt{chrome.exe}, \texttt{brave.exe}, and \texttt{msedge.exe}. The process family name labels generalize the process name labels. We use precision, $tp/(tp+fp)$, and recall, $tp/(tp+fn)$, to highlight the system's performance on individual processes and malware. For the malware detection results, we use a binary label where samples are considered malware if five or more anti-virus engines labeled the SHA-256 associated with the process as malicious.

\bgroup
\def\arraystretch{1.1}
\begin{table}
\center
\begin{tabular}{lrr}
  \hline
  \multicolumn{1}{c}{Method} & \multicolumn{1}{c}{Process Family} & \multicolumn{1}{c}{Process} \\
                             & \multicolumn{1}{c}{$F_1$ Score}       & \multicolumn{1}{c}{$F_1$ Score} \\
  \hline
  W-Na\"{i}ve Bayes     & 0.9941 (+/- 0.0004) & 0.9650 (+/- 0.0013) \\
  Na\"{i}ve Bayes       & 0.9879 (+/- 0.0007) & 0.9571 (+/- 0.0012) \\
  Top Process           & 0.8953 (+/- 0.0043) & 0.8860 (+/- 0.0052) \\
  \texttt{server\_name} & 0.8556 (+/- 0.0073) & 0.8215 (+/- 0.0068) \\
  \texttt{dst\_ip}      & 0.8537 (+/- 0.0154) & 0.8181 (+/- 0.0154) \\
  \hline
\end{tabular}
\caption{Process inference results on data collected from Site 1 (GMT+19). The \texttt{server\_name} and \texttt{dst\_ip} methods do not use the fingerprint information but rely strictly on the specified destination information.}
\label{table:proc-inference-accuracy}
\end{table}
\egroup

All results in this section use a knowledge base constructed on data collected during May 2020 from Site 1 (GMT+19) and the malware analysis sandbox. As we show in Section \ref{sec:operationalizing}, including the training data from the months prior to May does not increase the performance of the system. The testing data is collected during the first 10 days of June 2020 from Site 1 (GMT+19), Site 2 (GMT+8), and the malware analysis sandbox as specified throughout this section.

\subsection{Process Identification}

The core feature of the system described in Section \ref{sec:generalized-tls-fingerprinting} is to infer the process name from the TLS fingerprint string and destination information. Table \ref{table:proc-inference-accuracy} lists an overview of these results when applied to the first ten days of data collected in June 2020 from Site 1 (GMT+19). The $F_1$ score is averaged over each day and presented with its standard deviation. The baseline method of selecting the process with the most observations in the knowledge base for a given fingerprint string resulted in an $F_1$ score of 0.8953. Both the unweighted and weighted na\"{i}ve Bayes methods improved significantly on the baseline, with the weighted na\"{i}ve Bayes method achieving an $F_1$ score of 0.9941 for process family identification. A strategy that ignores the fingerprint information and selects the process most closely associated with either the TLS \texttt{server\_name} or destination IP address performed significantly worse. A diverse set of processes often communicate with the same set of destinations, and the TLS fingerprint string is needed to achieve superior process identification performance.

The sessions misclassified by the weighted na\"{i}ve Bayes algorithm were skewed towards a small set of misclassifications between Microsoft Outlook, Cisco Webex, and Safari. Cisco Webex has components that integrate with Microsoft Outlook, resulting in sessions initiated by Cisco Webex that communicate with \texttt{outlook.office365.com} on both WinNT and MacOS, confusing the classifier. There is an overlap in the fingerprint strings that Microsoft Outlook, Cisco Webex, and Safari present on MacOS. When Microsoft Outlook or Cisco Webex communicate with CDNs or advertising sites using the default CoreTLS library, both the fingerprint string and the destination are more strongly correlated with Safari in the knowledge base, resulting in misclassifications. The aforementioned cases account for $\sim$40\% of the misclassified sessions. The other major outlier is Chromium-based applications like Electron and Slack being misclassified as Chromium-based web browsers. This case is responsible for $\sim$20\% of the misclassifications.

\begin{figure}
	\centering
    \includegraphics[scale=0.50]{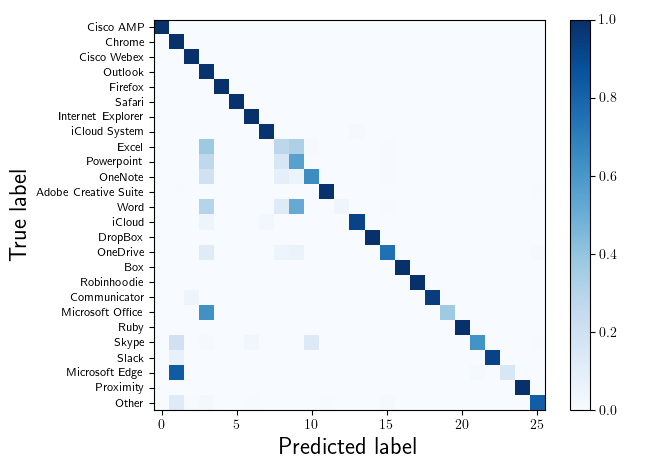}
	\caption{Confusion matrix for the top-25 processes on data collected from Site 1 (GMT+19).}
	\label{fig:top-process-confusion-matrix}
\end{figure}

\bgroup
\def\arraystretch{1.1}
\begin{table*}
\center
\begin{tabular}{|l|r|r|r||r|r||r|r||r|r|}
  \hline
  \multicolumn{1}{|c|}{\multirow{2}{*}{Process}} & \multicolumn{1}{c|}{\multirow{2}{*}{Sessions}} & \multicolumn{2}{c||}{W-Na\"{i}ve Bayes} & \multicolumn{2}{c||}{Top Process} & \multicolumn{2}{c||}{\texttt{server\_name}} & \multicolumn{2}{c|}{\texttt{dst\_ip}} \\
  \cline{3-10}
   &  & \multicolumn{1}{c|}{Precision} & \multicolumn{1}{c||}{Recall} & \multicolumn{1}{c|}{Precision} & \multicolumn{1}{c||}{Recall} & \multicolumn{1}{c|}{Precision} & \multicolumn{1}{c||}{Recall} & \multicolumn{1}{c|}{Precision} & \multicolumn{1}{c|}{Recall} \\
  \hline
  Cisco AMP            & 105,889,766 & 0.9999 & 1.0000 & 0.9851 & 0.9999 & 0.9576 & 0.9999 & 0.9997 & 0.9935 \\
  Chromium             & 45,931,514  & 0.9920 & 0.9998 & 0.9781 & 0.9993 & 0.6439 & 0.3059 & 0.6293 & 0.8496 \\
  Cisco Webex          & 38,403,011  & 0.9989 & 0.9923 & 0.8141 & 0.9680 & 0.9819 & 0.8972 & 0.9805 & 0.9653 \\
  Microsoft Office     & 26,855,990  & 0.9788 & 0.9911 & 0.7887 & 0.3963 & 0.9290 & 0.9745 & 0.9098 & 0.9647 \\
  Firefox              & 22,234,838  & 0.9994 & 0.9999 & 0.9992 & 0.9999 & 0.6329 & 0.3059 & 0.6124 & 0.2689 \\
  Safari               & 7,633,503   & 0.9787 & 0.9903 & 0.3751 & 0.9072 & 0.5094 & 0.2210 & 0.4672 & 0.1861 \\
  Internet Explorer    & 4,373,033   & 0.9903 & 0.9969 & 0.9490 & 0.8761 & 0.6521 & 0.2875 & 0.5343 & 0.2731 \\
  iCloud               & 4,328,783   & 0.9658 & 0.9803 & 0.6135 & 0.2546 & 0.9242 & 0.8512 & 0.8770 & 0.8230 \\
  Creative Cloud       & 1,891,238   & 0.9955 & 0.9950 & 0.5110 & 0.1246 & 0.9852 & 0.9881 & 0.9222 & 0.6824 \\
  Box                  & 664,518     & 0.9992 & 0.9961 & 0.9822 & 0.9239 & 0.9555 & 0.9992 & 0.9508 & 0.9072 \\
  \hline
\end{tabular}
\caption{Process inference results for the top-10 most prevalent process families on data collected from Site 1 (GMT+19).}
\label{table:proc-inference-accuracy-process}
\end{table*}
\egroup

Figure \ref{fig:top-process-confusion-matrix} and Table \ref{table:proc-inference-accuracy-process} present a more detailed view of the weighted na\"{i}ve Bayes classifier's performance on the most prevalent processes in the test data. Figure \ref{fig:top-process-confusion-matrix} presents the confusion matrix for the top-25 process names in the test data. The ``Other" category in the confusion matrix consists of all remaining processes. In general, we can correctly identify individual processes using the proposed methods. The primary weakness is disambiguating processes that share a common architecture and purpose. For example, Microsoft Office applications are often confused. The Chromium-based Microsoft Edge is also misclassified as Chrome. In terms of implementing a network security policy, process specificity may not be needed, and general process families may suffice. For example, a security policy could allow all Microsoft Office applications to bypass the firewall and communicate directly with Microsoft servers.

Table \ref{table:proc-inference-accuracy-process} lists the precision and recall for the 10 most prevalent process families. These process families all have precision and recall greater than 0.99 for the weighted na\"{i}ve Bayes classifier with three exceptions: Microsoft Office, Safari, and iCloud. As was the case for specific process categories, the lower performance on these process families is due to other applications integrating with underlying services and shared fingerprint strings communicating with generic CDNs. These top-10 process families accounted for 93.41\% of the total test traffic from Site 1 (GMT+19). If we remove the data associated with these top-10 families from the testing data, the weighted na\"{i}ve Bayes classifier still achieves an $F_1$ score of 0.9711, with over 60\% of the misclassifications due to Chromium-based applications being misclassified as Chromium-based web browsers. The performance of the classifiers based solely on destination information is significantly worse for browsers that go to a wide variety of destinations and generally underperforms on all processes.

\bgroup
\def\arraystretch{1.1}
\begin{table}
\center
\begin{tabular}{lrr}
  \hline
  \multicolumn{1}{c}{Method} & \multicolumn{1}{c}{Process Family}  & \multicolumn{1}{c}{Process} \\
                             & \multicolumn{1}{c}{$F_1$ Score}        & \multicolumn{1}{c}{$F_1$ Score} \\
  \hline
  W-Na\"{i}ve Bayes     & 0.9858 (+/- 0.0019) & 0.9702 (+/- 0.0021) \\
  Na\"{i}ve Bayes       & 0.9786 (+/- 0.0040) & 0.9599 (+/- 0.0038) \\
  Top Process           & 0.9077 (+/- 0.0057) & 0.9035 (+/- 0.0061) \\
  \texttt{server\_name} & 0.8381 (+/- 0.0108) & 0.8297 (+/- 0.0091) \\
  \texttt{dst\_ip}      & 0.8145 (+/- 0.0286) & 0.7904 (+/- 0.0271) \\
  \hline
\end{tabular}
\caption{Process inference results on data collected from Site 2 (GMT+8).}
\label{table:proc-inference-accuracy-geography}
\end{table}
\egroup

In order to assess how well the techniques generalize to new networks, we used the knowledge base trained on data from Site 1 (GMT+19) in May 2020 to test on data collected from Site 2 (GMT+8) during the first ten days of June 2020. The results are summarized in Table \ref{table:proc-inference-accuracy-geography}. Selecting the most prevalent process without considering destination information performed slightly better compared to the results presented in Table \ref{table:proc-inference-accuracy}, but this is simply because there was less process diversity on Site 2 (GMT+8).

The weighted na\"{i}ve Bayes method remained competitive with a process family $F_1$ score of 0.9858 (compared with .9941 on Site 1). The reduction in performance is in part due to observing processes not seen on Site 1. There were over 300 unknown processes initiating over 100,000 TLS sessions, or 0.3\% of the total number of TLS sessions collected from Site 2. The remaining discrepancy is explained by geographic specific sub-domains and IP addresses that did not appear in the original database, biasing the classifier to select processes with higher prior probabilities.

As we discuss in Section \ref{sec:discussion}, the knowledge base would ideally be conditioned on enterprise and geography-specific data before deployment. In cases where this isn't feasible, Table \ref{table:proc-inference-accuracy-geography} demonstrates that competitive performance is still possible. It remains an open question how well the knowledge base would translate to data collected from a distinct enterprise, which we leave for future work.

\subsection{Cloud Orchestration and Processing Tools}
\label{sec:pets}

\bgroup
\def\arraystretch{1.1}
\begin{table*}
\center
\begin{tabular}{|l|r|r|r||r|r||r|r||r|r|}
  \hline
  \multicolumn{1}{|c|}{\multirow{2}{*}{Process}} & \multicolumn{1}{c|}{\multirow{2}{*}{Sessions}} & \multicolumn{2}{c||}{W-Na\"{i}ve Bayes} & \multicolumn{2}{c||}{Top Process} & \multicolumn{2}{c||}{\texttt{server\_name}} & \multicolumn{2}{c|}{\texttt{dst\_ip}} \\
  \cline{3-10}
   &  & \multicolumn{1}{c|}{Precision} & \multicolumn{1}{c||}{Recall} & \multicolumn{1}{c|}{Precision} & \multicolumn{1}{c||}{Recall} & \multicolumn{1}{c|}{Precision} & \multicolumn{1}{c||}{Recall} & \multicolumn{1}{c|}{Precision} & \multicolumn{1}{c|}{Recall} \\
  \hline
  \hline
  terraform-provider-aws & 81,784 & 0.9634 & 0.9975 & 0.5116 & 1.0000 & 0.9418 & 0.9247 & 0.8459 & 0.7754 \\
  docker                 & 45,412 & 0.9979 & 0.9993 & 0.1882 & 0.3043 & 0.5008 & 0.9914 & 0.4708 & 0.8991 \\
  terraform              & 7,382  & 0.9849 & 0.6816 & 0.4038 & 0.0573 & 0.6607 & 0.6140 & 0.3120 & 0.2614 \\
  kubectl                & 5,653  & 0.9684 & 0.9858 & 0.8847 & 0.6474 & 0.9975 & 0.6215 & 0.9285 & 0.6282 \\
  helm3                  & 690    & 0.9568 & 0.7696 & 0.9649 & 0.6783 & 0.8500 & 0.0986 & 0.8599 & 0.2580 \\
  awskinesistap          & 425    & 1.0000 & 0.9642 & 0.0000 & 0.0000 & 0.8182 & 0.2084 & 0.4889 & 0.0463 \\
  \hline
\end{tabular}
\caption{Process inference results for various cloud orchestration and processing applications found in the data collected from Site 1 (GMT+19) during June 2020.}
\label{table:pet-proc-inference-accuracy-process}
\end{table*}
\egroup

Tools specifically designed to facilitate cloud orchestration and processing serve an important role in the current network ecosystem. Identifying and prioritizing network traffic associated with these tools can provide an enhanced user experience resulting in increased productivity. Given applications' reliance on cloud computing resources, simply relying on IP addresses and domain names to differentiate processes consuming cloud services versus those responsible for running business critical services is difficult.

Table \ref{table:pet-proc-inference-accuracy-process} provides the classification results for six different cloud and virtualization tools. For these results, we selected processes that facilitated cloud or virtualized workflows and also were represented in our datasets. Terraform \cite{terraform} generally manages infrastructure resources, with terraform-provider-aws explicitly exposing AWS APIs. AWS Kinesis \cite{awskinesis} is a streaming data processing platform. Docker \cite{docker}, Kubernetes \cite{kubectl}, and Helm \cite{helm} all support deploying and managing containerized applications.

As shown in Table \ref{table:pet-proc-inference-accuracy-process}, the weighted na\"{i}ve Bayes classifier outperformed all competing approaches. Our approach performed well with respect to precision and recall for most applications, despite many of these applications connecting to generic AWS services like S3, which highlights the importance of incorporating the TLS fingerprint string into the analysis.

\subsection{Malware Detection}
\label{sec:malware-results}

\bgroup
\def\arraystretch{1.1}
\begin{table}
\center
\begin{tabular}{lrr}
  \hline
  \multicolumn{1}{c}{Method} & \multicolumn{1}{c}{Precision}  & \multicolumn{1}{c}{Recall} \\
  \hline
  W-Na\"{i}ve Bayes     & 0.9993 & 0.8868 \\
  Na\"{i}ve Bayes       & 0.9992 & 0.7038 \\
  Top Process           & 0.9731 & 0.0682 \\
  \texttt{server\_name} & 0.7792 & 0.6949 \\
  \texttt{dst\_ip}      & 0.6359 & 0.6499 \\
  \hline
\end{tabular}
\caption{Malware detection results for Site 1 (GMT+19) and the malware analysis sandbox.}
\label{table:malware-results}
\end{table}
\egroup

Malware has been replacing HTTP with TLS over the past several years \cite{anderson16deciphering}, and we observed malware samples from the malware analysis sandbox more frequently using TLS compared to HTTP during our collection period. The top two benign domains visited were \texttt{twitter.com} (14.5\% of the connections) and \texttt{www.google.com} (1.8\% of the connections). The \texttt{server\_name} extension was present in 98.2\% of connections, higher than the 90\% observed at Site 1 (GMT+19) during the same time. For the testing data, a total of 12,249 unique \texttt{server\_name} values were observed.

Table \ref{table:malware-results} presents the malware detection precision and recall for the malware analysis sandbox data and data collected from Site 1 (GMT+19). While the method of simply selecting the most prevalent process had a relatively high precision of 97.31\%, the recall was poor. The top process method only detected 6.82\% of the malware connections. This is unsurprising due to malware's reliance on system-provided TLS libraries. In our dataset, over 90\% of the malicious connections used the default Windows Schannel library \cite{schannel}, which generates TLS fingerprint strings used by many popular Windows process such as Microsoft Office and Internet Explorer.

By leveraging the destination information contained within the knowledge base and the weighted na\"{i}ve Bayes algorithm, we increased the recall to 88.68\%. Malware-initiated connections to \texttt{*.baidu.com} and \texttt{*.googleusercontent.com} were responsible for $\sim$45\% of the false negatives. In these cases, the malware samples used an Schannel-generated TLS fingerprint string, and there were an overwhelming number of connections to those domains by benign processes. Due to malware's reliance on popular hosting services like \texttt{*.googleusercontent.com}, the performance of classifiers strictly looking at destination information underperformed with precisions and recalls between 63\% and 77\%. The added information from the TLS fingerprint string helps in disambiguating many of the sessions that would be misclassified by the destination information alone.

To evade detection, malware authors could shift to using the TLS libraries of popular processes such as Chrome to avoid a trivial detection through unique or rare TLS fingerprint strings. These libraries also offer the most flexibility in terms of potential destinations that would blend into previous observations from those libraries, e.g., CDNs. But, unlike the developers of benign applications, malware authors are under additional constraints such as avoiding any noticeable user experience differences on the infected machine and the potential for take-down requests impacting their server infrastructure. If malware were to mimic popular fingerprint strings, the authors would need to make frequent updates to ensure their selected fingerprint string is still relevant. Our system automatically incorporates the latest fingerprint string, process, and destination information so that we are as robust as possible to the changing TLS landscape and can incorporate the latest malware trends into the knowledge base.

\subsection{Approximate Matching}
\label{sec:approx-matching}

\bgroup
\def\arraystretch{1.1}
\begin{table}
\center
\begin{tabular}{lrr}
  \hline
  \multicolumn{1}{c}{Site} & \multicolumn{1}{c}{Process Family}  & \multicolumn{1}{c}{Process} \\
                           & \multicolumn{1}{c}{$F_1$ Score}             & \multicolumn{1}{c}{$F_1$ Score} \\
  \hline
  Site 1 (GMT+19)     & 0.9024 (+/- 0.0577) & 0.9003 (+/- 0.0582) \\
  Site 2 (GMT+8)      & 0.7897 (+/- 0.0843) & 0.7622 (+/- 0.0849) \\
  \hline
\end{tabular}
\caption{Process inference results when restricted to fingerprint strings not in the database. In these cases, the analysis algorithm must rely on approximate matching.}
\label{table:proc-inference-accuracy-approx}
\end{table}
\egroup

New TLS fingerprint strings are continuously introduced into the ecosystem, and a robust system needs to handle these cases. As described in Section \ref{sec:approx-matching-desc}, we use Levenshtein distance to find ``close" fingerprint strings, and then perform process identification on the close fingerprint string's process list. Similar to the process identification results using Site 2 (GMT+8), this method will fail if the process isn't in the knowledge base or the close fingerprint string's process list.

To understand the performance of approximate matching, we analyzed the results of the process identification system when the test data is restricted to only contain fingerprint strings \textbf{not} in the knowledge base. During the first 10 days of June 2020, there were 159 fingerprint strings that were not in the May 2020 knowledge base constructed from Site 1; there were 8,781 TLS sessions (out of 278,570,891 total sessions) generated from these fingerprint strings. We also analyze the data collected from Site 2, where there were 53 unknown fingerprint strings and 3,217 TLS sessions (out of 33,820,842 total sessions).

Table \ref{table:proc-inference-accuracy-approx} lists the process identification results for both sites. The system achieved an $F_1$ score of 0.9024 for the process family problem on the data from Site 1. The system had problems classifying processes not seen in the training data, as well as TLS scanners \cite{sslscan,sslyze} that exhibit a large diversity in TLS fingerprint strings. The process family $F_1$ score for Site 2 when restricted to fingerprint strings not in the knowledge base was 0.7897. The discrepancy between the results on the two sites is explained almost entirely by processes observed on Site 2 that were \textit{never} observed on Site 1. 

While the results for approximate matching are worse than results when there is an exact match, we believe the approximate matching technique provides a valuable addition to a TLS fingerprinting system. Additionally, with a well-curated knowledge base, the number of sessions requiring approximate matching should be low. Only 0.003\% and 0.010\% of sessions from Site 1 and Site 2 required approximate matching.

\subsection{Feature Importance}
\label{sec:feature-importance}

\bgroup
\def\arraystretch{1.1}
\begin{table}
\center
\begin{tabular}{lrr}
  \hline
  \multicolumn{1}{c}{Feature} & \multicolumn{1}{c}{Process Family}  & \multicolumn{1}{c}{Process} \\
  \multicolumn{1}{c}{Set}     & \multicolumn{1}{c}{$F_1$ Score}        & \multicolumn{1}{c}{$F_1$ Score} \\
  \hline
  fp, sni, ip, port & 0.9941 (+/- 0.0004) & 0.9650 (+/- 0.0013) \\
  fp, sni, ip       & 0.9940 (+/- 0.0004) & 0.9656 (+/- 0.0012) \\
  fp, sni, port     & 0.9876 (+/- 0.0006) & 0.9567 (+/- 0.0012) \\
  fp, ip, port      & 0.9811 (+/- 0.0039) & 0.9485 (+/- 0.0040) \\
  fp, sni           & 0.9938 (+/- 0.0004) & 0.9651 (+/- 0.0014) \\
  fp, ip            & 0.9885 (+/- 0.0039) & 0.9578 (+/- 0.0041) \\
  fp, port          & 0.8955 (+/- 0.0042) & 0.8862 (+/- 0.0051) \\
  fp                & 0.8953 (+/- 0.0043) & 0.8860 (+/- 0.0052) \\
  sni               & 0.8556 (+/- 0.0073) & 0.8215 (+/- 0.0068) \\
  ip                & 0.8537 (+/- 0.0154) & 0.8181 (+/- 0.0154) \\
  \hline
\end{tabular}
\caption{Process identification results when only considering subsets of the destination features using data collected from Site 1 (GMT+19). \texttt{server\_name} is represented as ``sni" and only using the fingerprint is represented as ``fp".}
\label{table:missing-features}
\end{table}
\egroup

The destination features have varying levels of information. With the weighted na\"{i}ve Bayes algorithm, we take the feature's importance into account through the weights listed in Table \ref{table:feature-weights}. But there are several ongoing initiatives to increase the privacy of TLS sessions by obfuscating destination information. For example, encrypting the entire ClientHello (ECHO) \cite{echo} is one approach to obfuscate the \texttt{server\_name} value. ECHO would result in all TLS sessions to a single service provider (CDN, cloud provider, etc.) offering the same \texttt{server\_name} value.

The \texttt{server\_name} is the destination feature with the highest weight in our system, and it is natural to question the efficacy of the system if that feature were to contain significantly less information. To better understand the importance of particular destination features in our system, we performed process identification using the weighted na\"{i}ve Bayes algorithm on Site 1's June 2020 data with different subsets of features. Table \ref{table:missing-features} provides results for each combination of feature sets, where the feature set includes the destination feature and related equivalence mappings. Table \ref{table:missing-features} also provides results when only destination information is utilized, which is denoted by the lack of the ``fp" identifier in the table.

For the weighted na\"{i}ve Bayes classifier, the first row in Table \ref{table:missing-features} with ``fp, sni, ip, port" is the performance when using all destination features and ``fp" is the performance when ignoring all destination features, which then defaults to selecting the most prevalent process. If it were the case that the \texttt{server\_name} extension was completely removed from all TLS sessions, our approach would still achieve an $F_1$ score of .9811 for process family identification. On the other hand, if IP addresses no longer contained the same amount of information, e.g., all servers were hosted on CloudFlare, the system can still achieve an $F_1$ score of .9938 with the \texttt{server\_name} value alone. The port features add little information when considering aggregate statistics like the $F_1$ score but do help in niche cases such as email and remote desktop application identification.

Considering evasion with respect to destination features and Tables \ref{table:feature-weights} and \ref{table:missing-features} provides some insights. Simply omitting the \texttt{server\_name} may not give the desired effects because this will alter the fingerprint string. For example, Psiphon \cite{psiphon} exhibits many different fingerprint strings, one of which attempts to imitate Chrome. In some cases, Psiphon imitates Chrome but omits the \texttt{server\_name}, which causes it to be identifiable from just its TLS fingerprint string. Robust evasion needs to jointly consider all destination features and the TLS fingerprint string, while at the same time making sure the destinations and fingerprint string remain prevalent in real-world traffic.

\section{Operationalizing}
\label{sec:operationalizing}

In the previous section, all results use a knowledge base constructed from data collected during May 2020 to classify data collected during the first 10 days of June 2020. Maintaining an up-to-date knowledge base that captures the relevant real-world traffic statistics is critical to the success of our solution. In this section, we study how parameters of the knowledge base, such as its age, affect the performance of our classifier. In the previous section, we also took the most probable process returned by the classifier, ignoring the score. Here, we show the impact of considering the score on the classifier's performance and the amount of data discarded.

\subsection{Knowledge Base Age}

\begin{figure}
	\centering
    \includegraphics[scale=0.45]{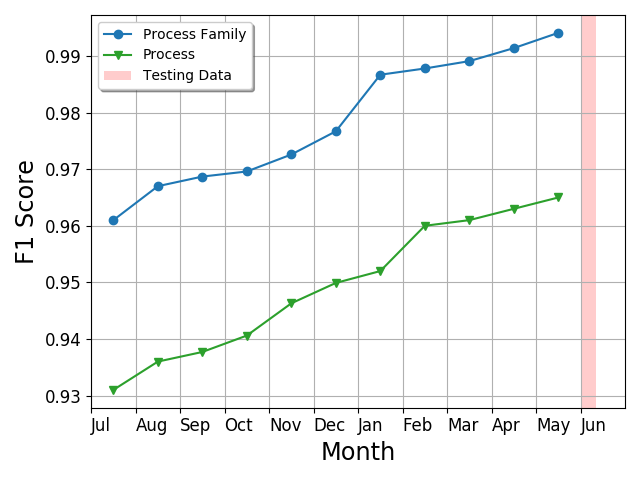}
	\caption{The effect of the knowledge base's age on classification performance, where the results use a knowledge base trained on the month specified by the x-axis.}
	\label{fig:database-age-effect}
\end{figure}

Maintaining the knowledge base with current, real-world data is not a trivial task, but is necessary due to the introduction of new fingerprint strings, processes, and destination information. We now investigate the importance of continuously updating the knowledge base by examining the performance of the system when only considering older data. For this experiment, we built 11 separate knowledge bases by merging daily knowledge bases for each month between July 2019 and May 2020. The May 2020 knowledge base was used in the experiments of the previous section.

Figure \ref{fig:database-age-effect} shows the process family and process classification results when the classifier only has access to data from the specified month, where the y-axis is from 0.93 to 0.99. Similar to the previous section, the testing data for each result is taken from the first 10 days of June 2020. As expected, the performance of the classifier on both label sets consistently decreases as the knowledge base is trained on older months.

The decreasing performance is due to the evolution of fingerprint strings and destination information. For example, while the testing data only contained 159 fingerprint strings with 8,781 TLS sessions that did not have a match in the knowledge base from May 2020, there were 1,238 fingerprint strings with 33 million TLS sessions in the testing data without a match in the July 2019 knowledge base. A heavier reliance on approximate matching will reduce the performance of the classifier as explained in Section \ref{sec:approx-matching}.

Even with approximate matching, Figure \ref{fig:database-age-effect} illustrates the clear need to keep an up-to-date knowledge base conditioned on recent real-world data as described in Section \ref{sec:knowledge-base}.

\subsection{Aging Out Data}

\begin{figure}
	\centering
    \includegraphics[scale=0.45]{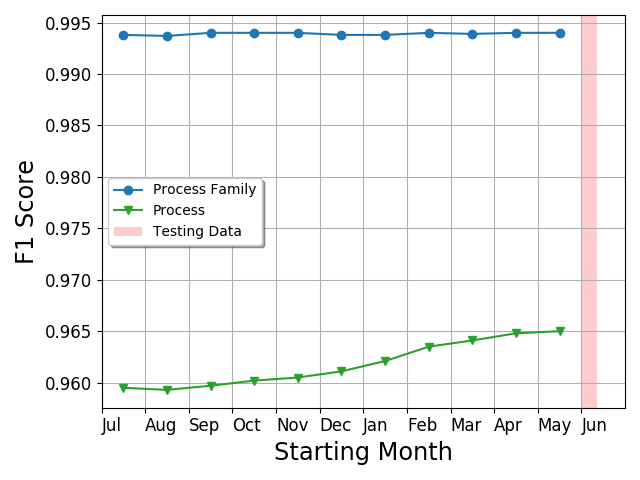}
	\caption{The effect of aging out older observations in the knowledge base on classification performance, where the results use a knowledge base constructed from data spanning the month specified by the x-axis until May 2020.}
	\label{fig:database-length-effect}
\end{figure}

In addition to keeping the knowledge base up to date with the most recent observations, one must consider the effect of older data that may no longer be relevant. Removing processes associated with a fingerprint string that have not been recently observed slightly increases performance and leads to a considerable reduction in the knowledge base's size.

To measure the impact of older data on our system, we again constructed 11 separate knowledge bases. For the current experiment, each knowledge base includes data starting from each of the months between July 2019 and May 2020. The 11 knowledge bases include all data between their starting month and the end of May 2020. The size of the knowledge bases became progressively smaller as we considered less data. The size of the knowledge constructed from data between July 2019 and May 2020 was 195 megabytes. The size of the knowledge base only considering the May 2020 data was 58 megabytes.

Figure \ref{fig:database-length-effect} illustrates the performance of the classifier when conditioned on each of the knowledge bases, using the first 10 days of June 2020 as testing data. There was no advantage in maintaining data older than one month with respect to process family and process classification $F_1$ scores. In fact, the performance of the process classification system decreased as the knowledge base kept data for longer periods of time. The decrease in performance was driven primarily by Chromium-based browsers being misclassified as Chrome due to lagging updates of BoringSSL \cite{boringssl}.

\subsection{Classifier Threshold}
\label{sec:classifier-threshold}

\begin{figure}
	\centering
    \includegraphics[scale=0.45]{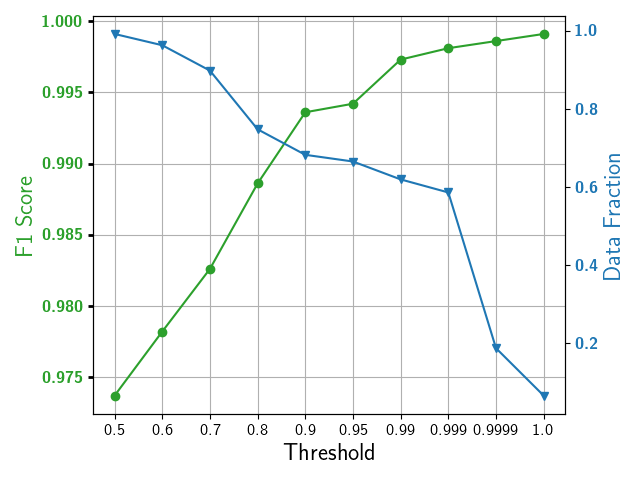}
	\caption{Effect of adjusting the classifier's threshold with minimum probability values in [0.5, 0.6, 0.7, 0.8, 0.9, 0.95, 0.99, 0.999, 0.9999, 1.0].}
	\label{fig:classifier-threshold-effect}
\end{figure}

The classification algorithm described in Section \ref{sec:naive-bayes} returns the most probable process associated with a fingerprint string and destination along with its probability. If network operators are unwilling or unable to accept misclassifications, they can define a minimum probability threshold and ignore any inferences that do not meet that threshold.

In this experiment, we used the May 2020 knowledge base and the June 2020 testing data from Site 1. Figure \ref{fig:classifier-threshold-effect} shows the impact of adjusting the classifier's threshold on the process inference $F_1$ score and the fraction of data discarded if we ignore results below the given threshold. For a threshold of 0.5, we classify over 99\% of the data and have a process inference $F_1$ score of 0.9737. At a threshold of 0.999, we classify 58\% of the data with an $F_1$ score of 0.9981. Finally, at a threshold of 1.0, we classify 6\% of the data with an $F_1$ score of 0.9991.

As the threshold is increased, the system begins to ignore (fingerprint string, destination) tuples that are observed with many processes. At the 0.999 threshold, the system ignores all off-diagonal sessions in Figure \ref{fig:top-process-confusion-matrix} except for $\sim$60\% of the Microsoft Edge connections because of Chrome's dominance with respect to its number of observations.

\section{Reproducibility}
\label{sec:reproducibility}

\begin{figure}
	\centering
    \includegraphics[scale=0.45]{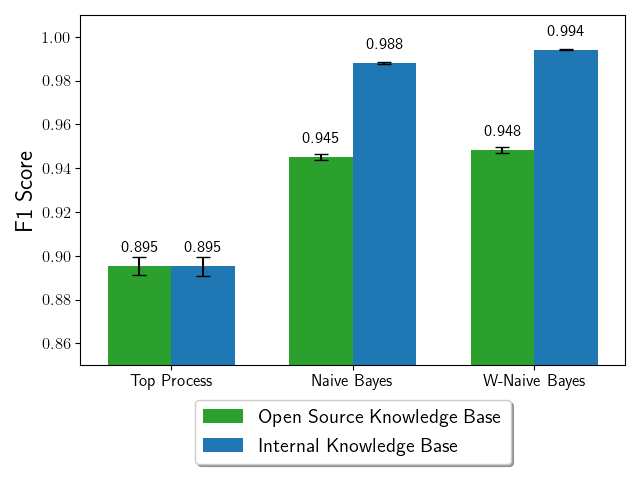}
	\caption{The difference in process family classification when using the open source and internal knowledge bases to test data collected from Site 1 (GMT+19) during June 2020.}
	\label{fig:public-vs-private-database}
\end{figure}

To assist reproducibility, we have open sourced the data collection and analysis system described in this paper. Our core tool is a C/C++ program that uses Linux's AF\_PACKET TPACKETv3 zero-copy shared memory ring buffers to collect and analyze data on network links with capacities of 30Gbps+. This tool supports the generation of TLS fingerprint strings and process inference with destination context as described in Sections \ref{sec:tls-fingerprinting} and \ref{sec:generalized-tls-fingerprinting}. mercury \cite{mercury} additionally generates client fingerprint strings for DHCP, DTLS, HTTP, SSH, and TCP, as well as server fingerprint strings for HTTP, DTLS, and TLS. We also released a pip-installable python implementation to facilitate rapid prototyping.

We released an open source version of our internal TLS fingerprint knowledge base along with the open source tools. We are committed to releasing up-to-date TLS fingerprint knowledge bases to the open source community, which we have currently done each week during the first seven months of 2020. The current open source knowledge base contains 8,405 unique TLS fingerprint strings with associated prevalence, process, and destination information. This dataset was constructed by considering 9.5 billion TLS sessions. There are currently 2,695 unique process names and 11,600 unique process executable SHA-256s in the open source knowledge base.

To comply with the policies of the organization whose sites we monitored, we had to remove some of the knowledge base's content. In the open source knowledge base, we only report the top-10 most prevalent processes per fingerprint string. For each process, we report only the equivalence mappings for the destination features. For example, the FQDN in the \texttt{server\_name} data is only reported as a domain name and TLD. There is a 30-day delay before observations on the monitored sites are introduced into the open source knowledge base. Finally, we did not open source any data from the malware analysis sandbox.

To better understand the impact of omitting data from the open source knowledge base, we compared the performance of the open source and internal knowledge bases when applied to data collected during the first 10 days of June 2020 from Site 1. We used the May 2020 internal knowledge base and the open source knowledge base that was available on June 1st, 2020. Figure \ref{fig:public-vs-private-database} illustrates the difference in performance when only selecting the most prevalent process and applying both the unweighted and weighted na\"{i}ve Bayes algorithms.

As expected, the difference in performance between the two knowledge bases was small when relying on the most prevalent process. The algorithms based on na\"{i}ve Bayes had a significant advantage when using the private knowledge base. As Figure \ref{fig:database-age-effect} demonstrated, the 30-day delay had a small impact on performance. Removing the most informative destination features, the \texttt{server\_name} and IP address according to Table \ref{table:feature-weights}, was the primary cause for the degraded performance. While the process inference performance based on the open source knowledge base is somewhat reduced, we still believe this data represents a significant contribution to the community and the first to associate TLS fingerprint strings, processes, and destination information on a large-scale.

\section{Related Work}
\label{sec:background}

The work presented in this paper builds on a rich history of network traffic fingerprinting and analysis. TLS fingerprinting first became popular in 2009 when Ivan Risti\'{c} released an Apache module to monitor SSL handshakes and correlate offered cipher suite lists with HTTP \texttt{User-Agent} strings \cite{ivan09modssf}. This led to several open source packages that implemented methods to extract TLS fingerprint strings and provided TLS fingerprint databases \cite{ivan2012git,p0f12ssl,brotherston15ftls,salesforce17ja}. The previous fingerprint databases did not provide real-world prevalence or contextual information about the destinations and therefore could not rely on that information to disambiguate the set of processes that mapped to the same fingerprint string. Hus\'{a}k et al. \cite{husak2015network} provided the first academic study of TLS fingerprinting, but again, did not consider destination information or have the infrastructure in place to develop detailed knowledge bases.

While our goal was to perform process inference using TLS fingerprinting, several efforts have used TLS fingerprinting as a means to perform measurement studies \cite{anderson2019beyond,frolov2019censorship,holz2015tls,kotzias2018coming,razaghpanah2017studying}. For example, Kotzias et al. \cite{kotzias2018coming} used a combination of open source fingerprint databases and their own data to examine how popular browsers modified the cryptographic parameters offered in their \texttt{client\_hello}'s in response to the disclosure of high-profile attacks against TLS \cite{alfardan2013security,bhargavan2016practical}. Frolov and Wustrow \cite{frolov2019censorship} studied the uniqueness of censorship circumvention tools' TLS fingerprint strings to motivate the development uTLS \cite{utls}, a TLS library to mimic and randomize the \texttt{client\_hello}. Our work illustrates the importance of considering destination features for libraries like uTLS when constructing a \texttt{client\_hello}.

Performing protocol, application, and process identification on encrypted traffic \cite{anderson2017noisy,anderson16identifying,liberatore2006inferring,taylor16appscanner,van2020flowprint,bernaille2007early,DBLP:journals/ccr/CrottiDGS07,moore2005internet,wright2006inferring,zhang2018homonit} has been an active area of research over the past 15 years. Initial work focused on identifying the application layer protocols, e.g., FTP, HTTP, and SMTP, within an encrypted tunnel \cite{moore2005internet,wright2006inferring}. For example, Wright et al. \cite{wright2006inferring} used the sequence of TCP packets and a hidden Markov model to identify application layer protocols.

More recent work has focused on the mobile and IoT domains \cite{taylor16appscanner,van2020flowprint,zhang2018homonit}. FlowPrint \cite{van2020flowprint} takes a semi-supervised approach that allows it to fingerprint previously unseen mobile applications. FlowPrint considers timing, device, and destination features, where the optimal batch window was found to be 300 seconds. Our work differs is several key areas but is complimentarily. Our system provides process identification results after having only seen the first packet in a TLS session, as opposed to 300 seconds in the worst case. We use a continuous data collection system to ensure our knowledge base has the most recent information, but we are unable to identify previously unseen processes.


\section{Discussion}
\label{sec:discussion}

The system presented in this paper relies on the large-scale collection, curation, and fusion of real-world data. We were able to achieve our results by creating a custom tool to collect network data and leveraging data from a pre-existing host agent. This approach led to quick results, but also created some critical gaps in our system's coverage. As detailed in Section \ref{sec:data}, the endpoints that generated host logs were almost entirely MacOS and WinNT-based desktop systems. There was a small Linux component, but a complete absence of mobile and IoT devices. While future work will include expanding the capabilities of the data collection system to remove these blind spots, we believe the underlying system and process inference strategy is sound and can naturally incorporate this new data.

Our data being limited to a single enterprise was another byproduct of our data collection strategy. We believe that the diversity of processes in our knowledge base and the results when applying the classification system to Site 2 (GMT+8) provide some evidence that our approach would scale to networks operated by a distinct enterprise. But, for optimal performance, having a knowledge base that was at least in part conditioned on data observed from the target site would be best. Any site with standard endpoint visibility agents with similar capabilities to the AnyConnect Network Visibility Module \cite{anyconnect} and the capacity to perform network monitoring could create custom knowledge bases, but this does require a significant initial investment. Further experiments into understanding how well the knowledge base transfers between distinct enterprises is left for future work.

Evading a system that continuously learns from billions of real-world TLS connections is not trivial, but we provided some best practices that privacy enhancing technologies could employ in Section \ref{sec:malware-results}, e.g., using system-provided TLS libraries. On the other hand, it is possible for malware to use these same techniques to evade detection. We hypothesize that the additional constraints placed on many classes of malware, e.g., maintaining prolonged periods of not being detected, make evading a continuously updating knowledge base substantially more difficult. While techniques based on information extracted only from the TLS \texttt{client\_hello} are not incapable of being evading, the results of Section \ref{sec:malware-results} indicate that our system does have value. More investigations into the security-privacy tradeoff of our system with respect to privacy enhancing technology and malware detection is needed.

There exists several avenues to extend the core methods of Section \ref{sec:generalized-tls-fingerprinting}. The most straightforward extension is to expand the set of destination feature equivalence mappings. Obvious examples include the global popularity or a binned consonant-to-vowel ratio of the \texttt{server\_name}. Adding additional destination features may also improve the performance of the system. In the May 2020 data, only 25\% of the TLS fingerprint strings and 35\% of the TLS sessions signaled support for TLS 1.3, and TLS 1.2 will most likely remain a large fraction of the TLS traffic for years to come. For TLS 1.2 sessions, including features around the server's certificate will provide additional information about the server's identity to the classification system.

Finally, maintaining proper ethics when performing a project analyzing real-world network and host data is critical. The unprocessed data was stored on a platform with an institution approved access control system. The data in the knowledge base was stripped of any indicators that could be used to identify users, such as the source IP addresses, detailed timestamps, and host agent identifiers. We followed all institutional procedures, including signing institutional agreements declaring that we would ``minimize personally identifiable information, maintain the confidentiality of all raw and processed data, receive written consent from your direct management chain before releasing any data, and pledge to not follow any practices that could be deemed discriminatory".





\section{Conclusion}
\label{sec:conclusion}

In this paper, we presented a system that continuously collects and fuses billions of real-world TLS sessions and host logs to generate a knowledge base correlating TLS client fingerprint strings, host processes, and destinations features. With the generated knowledge base, we built a system that uses a weighted na\"{i}ve Bayes algorithm to infer processes and detect malware using only the TLS fingerprint string and destination information contained within the first data packet of a TLS session. We demonstrated that our system was able to achieve an $F_1$ score of over 0.99 when inferring the process family, and high efficacy malware detection with 99.9\% precision and 88.7\% recall. We additionally examined the performance of our system when used to identify cloud orchestration and processing tools and found that the precision and recall were greater than 0.99 for several popular processes belonging to this category.

To assist in reproducibility, we contributed mercury \cite{mercury} to the open source community for collecting and classifying network traffic. We also released an open source version of our internal TLS fingerprint knowledge base, which is updated weekly and is currently the largest and most informative open source TLS fingerprint knowledge base in existence.

\begin{acks}
We thank Brandon Enright for his support in developing mercury. We thank both Brandon and Adam Weller for their feedback and support. We thank Lucas Messenger, Eddie Allan Jr., and Joey Rosen for their assistance in maintaining and providing access to the data capture infrastructure. We also thank and acknowledge Ed Paradise for his ongoing support of this work.
\end{acks}

\bibliographystyle{ACM-Reference-Format}
\bibliography{tls_inferencing}

\appendix

\section{TLS Extensions with Data}
\label{app:tls-extensions}

\vspace{2mm}

\bgroup
\def\arraystretch{0.9}
\begin{center}
\begin{tabular}{lc}
  \hline
  \multicolumn{1}{c}{Extension Name} & \multicolumn{1}{c}{Extension Hex Code} \\
  \hline
  \texttt{max\_fragment\_length} & \texttt{0001} \\
  \texttt{status\_request} & \texttt{0005} \\
  \texttt{client\_authz} & \texttt{0007} \\
  \texttt{server\_authz} & \texttt{0008} \\
  \texttt{cert\_type} & \texttt{0009} \\
  \texttt{supported\_groups} & \texttt{000a} \\
  \texttt{ec\_point\_formats} & \texttt{000b} \\
  \texttt{signature\_algorithms} & \texttt{000d} \\
  \texttt{heartbeat} & \texttt{000f} \\
  \texttt{application\_layer\_} & \texttt{0010} \\
  \texttt{ protocol\_negotiation} &  \\
  \texttt{status\_request\_v2} & \texttt{0011} \\
  \texttt{client\_certificate\_type} & \texttt{0013} \\
  \texttt{server\_certificate\_type} & \texttt{0014} \\
  \texttt{token\_binding} & \texttt{0018} \\
  \texttt{compress\_certificate} & \texttt{001b} \\
  \texttt{record\_size\_limit} & \texttt{001c} \\
  \texttt{supported\_versions} & \texttt{002b} \\
  \texttt{psk\_key\_exchange\_modes} & \texttt{002d} \\
  \texttt{signature\_algorithms\_cert} & \texttt{0032} \\
  \texttt{channel\_id} & \texttt{5500} \\
  \texttt{GREASE} & \texttt{0a0a} \\
  \hline
\end{tabular}
\end{center}
\egroup

\end{document}